\documentclass[aps,showpacs,twocolumn,amsmath,amssymb,superscriptaddress,prl]{revtex4-1}

\usepackage[dvipdfmx]{graphicx}
\usepackage{color}

\usepackage{bm}
\usepackage{comment}
\usepackage{color} 

\begin{document}

\title{Isotropic and anisotropic spin-dependent transport in epitaxial Fe$_3$Si}

\author{Nozomi Soya}
\affiliation{Department of Applied Physics and Physico-Informatics, Keio University, Yokohama 223-8522, Japan}

\author{Michihiro Yamada}
\affiliation{Center for Spintronics Research Network, Osaka University, 1-3 Machikaneyama, Toyonaka 560-8531, Japan}
\affiliation{PRESTO, Japan Science and Technology Agency, 4-1-8 Honcho, Kawaguchi, Saitama 332-0012, Japan}
\affiliation{Advanced Research Laboratories, Tokyo City University, 1-28-1
Tamazutsumi, Tokyo 158-8557, Japan}

\author{Kohei Hamaya}
\affiliation{Center for Spintronics Research Network, Osaka University, 1-3 Machikaneyama, Toyonaka 560-8531, Japan}
\affiliation{Spintronics Research Network Division, Institute for Open and Transdisciplinary Research Initiatives, Osaka University, Yamadaoka 2-1, Suita, Osaka 565-0871, Japan}

\author{Kazuya Ando}
\email{ando@appi.keio.ac.jp}
\affiliation{Department of Applied Physics and Physico-Informatics, Keio University, Yokohama 223-8522, Japan}
\affiliation{Keio Institute of Pure and Applied Sciences, Keio University, Yokohama 223-8522, Japan}
\affiliation{Center for Spintronics Research Network, Keio University, Yokohama 223-8522, Japan}

\begin{abstract}
We investigate spin-dependent transport phenomena in epitaxially grown Fe$_3$Si films, focusing on the anisotropic magnetoresistance (AMR), planar Hall effect (PHE), anomalous Hall effect (AHE), and spin Hall effect (SHE). While the sign and magnitude of the AMR and PHE depend on the current orientation relative to the crystallographic axes, the AHE and SHE remain nearly independent of the current orientation. The anisotropic AMR and PHE are attributed to current and magnetization dependent local band properties, including band crossing/anticrossing at specific $k$ points.
In contrast, the isotropic AHE and SHE arise from the Berry curvature integrated over the entire Brillouin zone, which cancels local variations. These findings highlight the interplay between symmetry, band structure, and magnetization in the spin-dependent transport phenomena.
\end{abstract}

\maketitle

\section{I. Introduction}

Spin-dependent transport in ferromagnetic metals (FMs) forms the foundation of non-equilibrium spintronics phenomena and various spintronics applications~\cite{DIENY1994335, harder2016electrical, 1058782, JULLIERE1975225, chappert2010emergence}. Among these phenomena, one of the most fundamental is anisotropic magnetoresistance (AMR), where the longitudinal electrical resistivity of an FM depends on the relative angle between the charge current and the magnetization~\cite{1058782}. Since the discovery of the AMR over 160 years ago~\cite{thomson1857xix}, its origin has been the subject of extensive investigation~\cite{doi:10.1063/1.1729509, SMIT1951612, JAOUL197723, PhysRevB.10.4626, doi:10.1063/1.4796178}.

Although it is now widely accepted that the AMR arises from spin-dependent scattering and spin-orbit coupling, a comprehensive understanding of this phenomenon remains challenging, particularly in single-crystal FMs~\cite{PhysRevB.79.092406, PhysRevB.63.134432, PhysRevB.84.094441, PhysRevB.77.205210, hupfauer2015emergence, PhysRevB.74.205205, doi:10.1063/1.4936175, doi:10.1063/1.352607, doi:10.1063/1.4927620, doi:10.1063/5.0034232, PhysRevLett.125.097201, HUNG2011372}.
 A recent work has demonstrated that the AMR of single-crystal Co$_x$Fe$_{1-x}$ films strongly depends on the current orientation with respect to the crystallographic axes~\cite{PhysRevLett.125.097201}. The current-orientation dependence of the AMR has been attributed to the anisotropy of the band structure at special $k$ points, where energy bands form crossing and anticrossing depending on the magnetization direction.
Furthermore, it has been demonstrated that even the sign of the AMR can reverse depending on the current orientation in single-crystal Fe$_3$Si films~\cite{HUNG2011372}. While the microscopic mechanism behind this behavior remains unclear, the anisotropic AMR in Fe$_3$Si films appears to reflect the intrinsic nature of its origin.


Spin-orbit coupling, the origin of AMR, also drives other spin-dependent transport phenomena, including the planar Hall effect (PHE), anomalous Hall effect (AHE), and spin Hall effect (SHE)~\cite{1058782, RevModPhys.82.1539, RevModPhys.87.1213}. An experimental study has shown that the intrinsic PHE in single-crystal Fe$_3$Si is closely connected to the AHE~\cite{friedland2006intrinsic}. This finding suggests that a comprehensive investigation of AMR, PHE, AHE, and SHE within a unified framework could enhance our understanding of the spin-dependent transport properties of single-crystal FMs.

In this paper, we investigate the current orientation dependence of the AMR, PHE, AHE, and SHE in epitaxial Fe$_3$Si films. Our results reveal that the sign and magnitude of the AMR and PHE vary with the current orientation relative to the crystallographic axes. In contrast, the sign and magnitude of the AHE and SHE remain unchanged regardless of the current orientation. These findings highlight the distinct origins of these spin-dependent transport phenomena, illustrating the unique role of current and magnetization dependent local band properties in determining their behavior.

\section{II. EXPERIMENTAL METHOD}

We conducted magnetoresistance and spin-orbit-torque (SOT) measurements to address the spin-dependent transport properties of epitaxial Fe$_3$Si. The magnetoresistance measurements were performed on epitaxially grown Fe$_3$Si devices with different current orientations. A 5-nm-thick epitaxial Fe$_3$Si(001) film with $D0_3$ ordering was grown on a single-crystal MgO(001) substrate by molecular beam epitaxy at a growth temperature below 80$^\circ$C~\cite{10.1063/1.2996581, PhysRevB.86.174406,PhysRevB.83.144411}. Then, a 4-nm-thick SiO$_2$ capping layer was deposited on the Fe$_3$Si film by magnetron sputtering at room temperature to prevent the oxidation of the Fe$_3$Si layer. The film was patterned into Hall bar shaped devices with a width of $20$~$\mu$m and a length of $200$~$\mu$m with the current orientation along [110], [010] and [$\bar{1}10$], as shown in Fig.~\ref{fig1}(a), by using the photolithography and Ar-ion milling. The current orientations of $\bf{J}$$\parallel [110]$ and $\bf{J}$$\parallel [\bar{1}10]$ are symmetrically identical due to the four fold crystalline symmetry of Fe$_3$Si.

For the investigation of the SHE in epitaxial Fe$_3$Si, we measured current-induced SOTs using the spin-torque ferromagnetic resonance (ST-FMR) for SiO$_2$(4~nm)/Co(5~nm)/Ti(3~nm)/Fe$_3$Si(5~nm)/MgO(001)-substrate devices, where the numbers in parentheses represent the thickness. The SiO$_2$/Co/Ti film was fabricated on the Fe$_3$Si(001) film by the magnetron sputtering at room temperature. The Co/Ti/Fe$_3$Si(001) film was patterned into rectangular strips with a width of $10$~$\mu$m and a length of $70$~$\mu$m by using the photolithography and Ar-ion milling. 
In the Co/Ti/Fe$_3$Si(001) devices, the in-plane magnetized Co layer is magnetically separated from the in-plane magnetized Fe$_3$Si layer by the Ti layer.

For the ST-FMR measurement, we applied a radio frequency (RF) current with a frequency of $f$ along the longitudinal direction of the device and swept an in-plane external field $H$ applied at an angle of $\theta_{xy}$ from the longitudinal direction. The applied RF current generates a spin current by the SHE in the Fe$_3$Si layer. The spin current is injected into the Co layer through the Ti layer with sufficiently long spin diffusion length~\cite{PhysRevB.90.140407}, exerting SOTs on the magnetization in the Co layer. Note that the AHE in the Fe$_3$Si layer does not exerts a torque on the magnetization in the Co layer because the magnetization in the Co layer and that in the Fe$_3$Si layer are aligned parallel by external magnetic field. 
The SOTs, as well as an Oersted field, drive the precession of the magnetization in the Co layer at the FMR field of the Co layer, $H=H_\mathrm{FMR,Co}$. The magnetization precession yields resistance oscillations due to the AMR of the Co layer. The change in the resistance mixes with the alternating current to create a direct current (DC) voltage $V_{\rm{DC, Co}}$ across the bar at $H=H_\mathrm{FMR,Co}$. 
The magnetization precession in the Fe$_3$Si layer also produces DC voltage $V_{\rm{DC, Fe_3Si}}$ through the AMR of the Fe$_3$Si layer at the FMR field of the Fe$_3$Si layer, $H=H_\mathrm{FMR,Fe_{3}Si}$. 
We measured $V_{\rm{DC}}=V_{\rm{DC, Co}}+V_{\rm{DC, Fe_3Si}}$ using a bias tee at room temperature.

\section{III. RESULTS AND DISCUSSION}

\subsection{A. Anisotropic magnetoresistance, planar Hall effect, and anomalous Hall effect}

\begin{figure}[tb]
\center\includegraphics[scale=1]{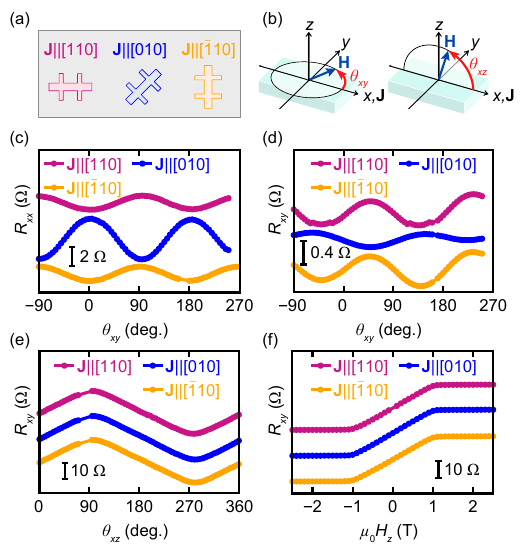}
	\caption{(a) A schematic illustration of the SiO$_2$/Fe$_3$Si/MgO(001)-substrate devices with different current orientations. (b) The definitions of the rotating angles $\theta_{xy}$ and $\theta_{xz}$ of the magnetic field. (c), (d), (e) The results of the field-angle-dependent magnetoresistance measurements for the Fe$_3$Si films with different current orientations: (c) $\theta_{xy}$ dependence of the longitudinal resistance $R_{xx}$; (d)  $\theta_{xy}$ dependence of the transverse resistance $R_{xy}$; (e) $\theta_{xz}$ dependence of $R_{xy}$. The measurements were done by rotating samples in a magnetic field $\mu_0 H=0.5~\rm T$ for $\theta_{xy}$-rotation and $\mu_0 H=3~\rm T$ for $\theta_{xz}$-rotation. (f) $R_{xy}$ as a function of the magnetic field $\mu_0 H_{z}$ applied along the $z$ axis for the Fe$_3$Si films with different current orientations.
}
\label{fig1} 
\end{figure}

First, we investigate the AMR and PHE of the Fe$_3$Si films by measuring the longitudinal resistance $R_{xx}$ and transverse resistance $R_{xy}$, respectively, with applying a magnetic field $\mu_{0} H$ = 0.5~T rotated in the $xy$ plane. The definitions of the angles of the magnetic field are shown in  Fig.~\ref{fig1}(b). When an external magnetic field rotates the magnetization of a FM, $R_{xx}$ and $R_{xy}$ in a single domain isotropic system are given by~\cite{1058782}
\begin{align}
R_{xx}& = R_{\perp}+(R_{\parallel}-R_{\perp}){\rm cos}^2\theta_{xy}\label{AMR}\\
R_{xy}& = \frac{1}{2} (R_{\parallel}-R_{\perp}){\rm sin}2\theta_{xy},\label{PHE}
\end{align}
where $R_{\perp}$ and $R_{\parallel}$ are the resistances when the magnetization is perpendicular and parallel to the current, respectively. Typical FMs, such as Fe, Co and Ni, exhibit a positive ($R_{\parallel}-R_{\perp}>0$) magnetoresistance, while some other materials, such as the half-metals, show a negative ($R_{\parallel}-R_{\perp}<0$) magnetoresistance~\cite{1058782, kokado2012anisotropic}.
Figure~\ref{fig1}(c) shows the $\theta_{xy}$ dependence of $R_{xx}$ for the Fe$_3$Si film. This result shows that the sign and amplitude of the AMR in the epitaxial Fe$_3$Si film depend on the current orientation; the magnetoresistance is negative for $\bf{J}$$\parallel [110]$ and $\bf{J}$$\parallel [\bar{1}10]$, while it is positive for $\bf{J}$$\parallel [010]$. This current orientation dependence of the AMR is consistent with a previous report on single-crystal Fe$_3$Si films~\cite{HUNG2011372}. The observed anisotropic AMR is reminiscent of the intrinsic mechanism of the AMR arising from the energy band crossing dependent on the magnetization direction, discovered in single-crystal Co$_x$Fe$_{1-x}$ films~\cite{PhysRevLett.125.097201}.

The sign and amplitude of the PHE also depend on the current orientation, as shown in Fig.~\ref{fig1}(d). Figures~\ref{fig1}(c) and \ref{fig1}(d) demonstrate that the sign of the PHE is opposite to that of the AMR across all current orientations, which contradicts the AMR-related origin of the PHE described by Eqs.~(\ref{AMR}) and (\ref{PHE}). This discrepancy has been observed in stoichiometric single-crystal Fe$_3$Si films with $\bf{J} \parallel [110]$ or $\bf{J} \parallel [\bar{1}10]$ grown on GaAs(001) substrates~\cite{friedland2006intrinsic, PhysRevB.71.172401}. Beyond the AMR-related origin described by Eq.~(\ref{PHE}), a PHE with an opposite sign has been attributed to intrinsic mechanisms involving the Berry phase and spin chirality~\cite{friedland2006intrinsic}. The observation that the sign of the PHE differs from that of the AMR indicates that the PHE in the epitaxial Fe$_3$Si film is dominated by the intrinsic mechanism rather than the AMR-based mechanism.

In contrast to the AMR and PHE, the sign and magnitude of the AHE are independent of the current orientation. Figure~\ref{fig1}(e) shows $R_{xy}$ measured with an applied magnetic field ($\mu_{0} H$ = 3~T) rotated in the $xz$ plane. The result demonstrates that all three devices exhibit nearly identical signals, $R_{xy} \sim +\sin \theta_{xz}$, arising from the positive AHE. Here, at $\mu_{0} H$ = 3~T, the magnetization of the Fe$_3$Si layer, which has in-plane magnetic anisotropy, is saturated and aligns with the external field direction, as confirmed by the out-of-plane magnetic field dependence of $R_{xy}$ shown in Fig.~\ref{fig1}(f). These results highlight a key difference: while the signs of the AMR and PHE vary with the current orientation, the sign of the AHE is independent of the current orientation. A summary of the current orientation dependence of the AMR, PHE, and AHE in the epitaxial Fe$_3$Si is presented in Table~\ref{table}. The results for $\mathbf{J} \parallel [010]$ and $\mathbf{J} \parallel [110]$ are shown; those for $\mathbf{J} \parallel [\bar{1}10]$ are identical to $\mathbf{J} \parallel [110]$, consistent with the fourfold crystalline symmetry of Fe$_3$Si.

\begin{table}[tb]
 \caption{Current orientation dependence of the sign of the AMR, PHE, AHE, and SHE in the single-crystal Fe$_3$Si films.}
 \label{table}
 \centering                 
  \begin{tabular}{cccccc}
   \hline \hline
    & AMR & PHE & AHE & SHE \\
   \hline
   $\bf{J}$$\parallel[010]$ & + & -- &+ & +\\
   $\bf{J}$$\parallel [110]$ & -- & + &+ & +\\
         \hline \hline
  \end{tabular}
\end{table}

\subsection{B. Spin Hall effect}

\begin{figure}[tb]
\center\includegraphics[scale=1]{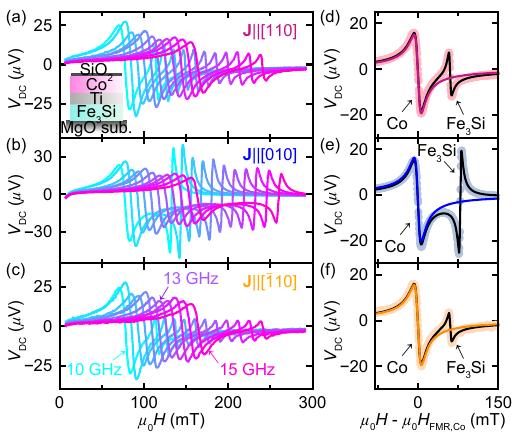}
\caption{The schematic illustration of the SiO$_2$/Co/Ti/Fe$_3$Si/MgO(001)-substrate device for the ST-FMR measurements is shown in the inset to (a). Magnetic field $\mu_{0} H$ dependence of the DC voltage $V_{\rm{DC}}$ for the Co/Ti/Fe$_3$Si films with (a) $\bf{J}$$\parallel [110]$, (b) $\bf{J}$$\parallel [010]$, and (c) $\bf{J}$$\parallel [\bar{1}10]$. The RF frequency was varied from $f$ = 10 to 15~GHz. $V_{\rm{DC}}$ spectra at $f$ = 13~GHz for the Co/Ti/Fe$_3$Si films with (d) $\bf{J}$$\parallel [110]$, (e) $\bf{J}$$\parallel [010]$, and (f) $\bf{J}$$\parallel [\bar{1}10]$. The spectra consist of the FMR peaks of the Co layer (left) and Fe$_3$Si layer (right) as indicated by the black arrows. The gray solid curves are the fitting results using Eq.~(\ref{Lolentz}). The green curves are the symmetric component and the pink curves are the anitisymmetric component of $V_{\rm{DC, Co}}$ due to the FMR of the Co layer. }
\label{fig2} 
\end{figure}
\begin{figure*}[tb]
\center\includegraphics[scale=1]{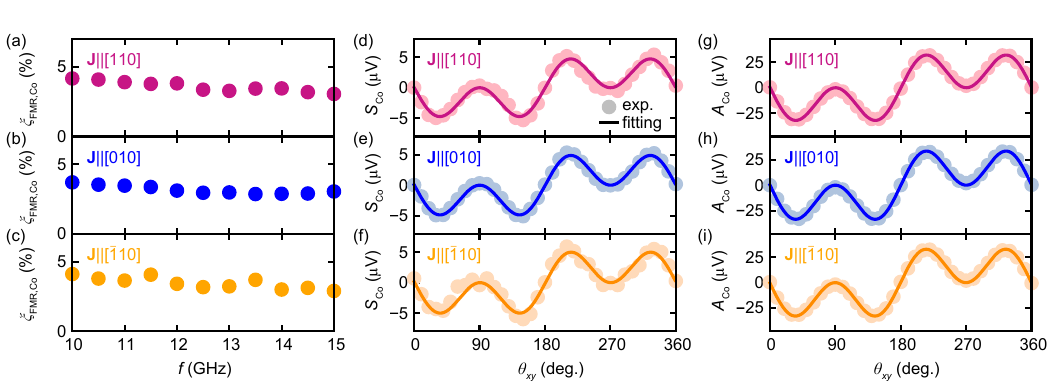}
\caption{ The RF frequency $f$ dependence of the FMR spin-torque generation efficiency $\xi_{\rm{FMR, Co}}$ from $f$ = 10 to 15~GHz for the Co/Ti/Fe$_3$Si(001) devices with (a) $\bf{J}$$\parallel [110]$, (b) $\bf{J}$$\parallel [010]$, and (c) $\bf{J}$$\parallel [\bar{1}10]$.
Magnetic field angle $\theta_{xy}$ dependence of the (d), (e), (f) symmetric components $S_{\rm{Co}}$ and (g), (h), (i) antisymmetric components $A_{\rm{Co}}$ of the ST-FMR spectra at $f = 13~\rm{GHz}$ for the Co/Ti/Fe$_3$Si(001) devices with (d), (g) $\bf{J}$$\parallel [110]$, (e), (h) $\bf{J}$$\parallel [010]$, and (f), (i) $\bf{J}$$\parallel [\bar{1}10]$. The solid circles are the experimental data, and the solid curve is the fitting result using a function proportional to sin2$\theta_{xy}$cos$\theta_{xy}$. 
}
\label{fig3} 
\end{figure*}

Next, we investigate the SHE in the epitaxial Fe$_3$Si using the ST-FMR. 
Figures~\ref{fig2}(a), \ref{fig2}(b), and \ref{fig2}(c) show the ST-FMR spectra measured at $\theta_{xy}=45^\circ$ for the Co/Ti/Fe$_3$Si devices with $\bf{J}$$\parallel[110]$, $\bf{J}$$\parallel[010]$, and $\bf{J}$$\parallel[\bar{1}10]$, respectively. This result shows that the resonance fields change systematically with $f$, consistent with the prediction of the ST-FMR.
 The ST-FMR spectra are composed of two peaks due to the FMR of the Co and Fe$_3$Si layers, as shown in Figs.~\ref{fig2}(d), \ref{fig2}(e), and \ref{fig2}(f): $V_{\rm{DC}}= V_{\rm{DC,Co}}+V_{\rm{DC,Fe_3Si}}$. 
Since the saturation magnetization, measured by using a vibrating sample magnetometer, of Co ($\mu_{0} M_{\rm{s, Co}}$ = 1.36~T) is larger than that of Fe$_3$Si ($\mu_{0} M_{\rm{s, Fe_3 Si}}$ = 1.08~T), the Kittel formula predicts that the peaks with the smaller and larger resonance fields correspond to the FMR of the Co and Fe$_3$Si layers, respectively~\cite{PhysRev.73.155}. 
In fact, the sign and magnitude of the ST-FMR signals with the larger resonance fields depend on the current orientation, consistent with the sign and magnitude of the AMR of the Fe$_3$Si layer observed in the previous section.

Each peak of the ST-FMR spectra consists of symmetric and and antisymmetric components as~\cite{PhysRevLett.106.036601}
\begin{widetext}
	\begin{equation}
V_{\rm{DC}}= \sum_{\rm{FM = Co, Fe_3 Si}} 
S_{\rm{FM}}\frac{W_{\rm{FM}}^2}{(\mu_0H-\mu_0H_{\rm{FMR, FM}})^2+W_{\rm{FM}}^2}
+A_{\rm{FM}}\frac{W_{\rm{FM}}(\mu_0H-\mu_0H_{\rm{FMR, FM}})}{(\mu_0H-\mu_0H_{\rm{FMR, FM}})^2+W_{\rm{FM}}^2},
\label{Lolentz}
\end{equation}
\end{widetext}
where $H_{\rm{FMR, FM}}$ is the resonance field of the FM(= Co, Fe$_3$Si) layer and $W_{\rm{FM}}$ is the spectral width. The symmetric component $S_{\rm{FM}}$ corresponds to the out-of-plane damping-like spin-orbit effective field, and the antisymmetric component $A_{\rm{FM}}$ corresponds to the in-plane field due to the Oersted field and field-like spin-orbit effective field. 
We fitted the measured $V_{\rm{DC}}$ signals using Eq.~(\ref{Lolentz}) as shown 
in Figs.~\ref{fig2}(d), \ref{fig2}(e) and \ref{fig2}(f), where the the gray solid curves are the total fitting result while the green(pink) solid curves represent the symmetric(antisymmetric) component of the ST-FMR signals $V_{\rm{DC, Co}}$ at the FMR field of the Co layer.

\begin{figure}[tb]
\center\includegraphics[scale=1]{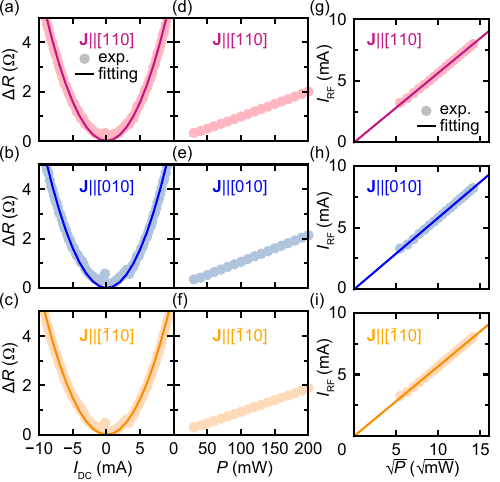}
\caption{DC current $I_{\rm DC}$ dependence of the resistance change $\Delta R(I_{\rm DC}) = R(I_{\rm DC}) - R(I_{\rm DC}=0)$ for the Co/Ti/Fe$_3$Si(001) device with (a) $\bf{J}$$\parallel [110]$, (b) $\bf{J}$$\parallel [010]$, and (c) $\bf{J}$$\parallel [\bar{1}10]$. The solid circles are the experimental result and the solid curve is the fitting result using a parabolic function. RF power $P$ dependence of the resistance change $\Delta R(P) = R(P) - R(P = 0)$ for the Co/Ti/Fe$_3$Si(001) device with (d) $\bf{J}$$\parallel [110]$, (e) $\bf{J}$$\parallel [010]$, and (f) $\bf{J}$$\parallel [\bar{1}10]$ (orange). The RF current $I_{\rm RF}$ for the Co/Ti/Fe$_3$Si(001) device with (g) $\bf{J}$$\parallel [110]$, (h) $\bf{J}$$\parallel [010]$, and (i) $\bf{J}$$\parallel [\bar{1}10]$ obtained from the heating calibration as a function of the square root of $P$. The solid circles are the experimental data, and the solid lines are the linear fit to the data.}
\label{fig4} 
\end{figure}

To investigate the SHE in the Fe$_3$Si layer, we characterize $V_{\rm{DC, Co}}$, where the FMR in the Co layer is driven by the SOTs due to the SHE in the Fe$_3$Si layer. 
Figures~\ref{fig3}(a), \ref{fig3}(b) and \ref{fig3}(c) show frequency $f$ dependence of the FMR spin-torque generation efficiency as~\cite{doi:10.1126/sciadv.aax4278}
\begin{equation}
\xi_{\mathrm{FMR, Co}}=\frac{S_{\rm{Co}}}{A_{\rm{Co}}} \frac{e \mu_{0} M_{\rm{s, Co}} d_{\rm{Co}} (d_{\mathrm{Ti/Fe_3Si}}^{\rm eff})}{\hbar} \sqrt{1+ \frac{M_{\rm{eff, Co}}}{ H_{\mathrm{FMR, Co}}}},
\end{equation} 
where $d_{\rm{Co}}$ and $M_{\rm{eff, Co}}$ are the thickness and effective demagnetization field of the Co layer, respectively. Here, $d_{\mathrm{Ti/Fe_3Si}}^{\rm eff} = d_{\mathrm{Fe_3Si}}+ (\rho_{\mathrm{Fe_3Si}}/\rho_{\mathrm{Ti}})d_{\mathrm{Ti}}$ is the effective thickness of the Ti/Fe$_3$Si bilayer, where $d_{\mathrm{Fe_3Si(Ti)}}$ and $\rho_{\mathrm{Fe_3Si(Ti)}}$ are the thickness and resistivity of the Fe$_3$Si(Ti) layer, respectively. 
Figures~\ref{fig3}(a), \ref{fig3}(b) and \ref{fig3}(c) show that $\xi_{\mathrm{FMR}}$ barely changes with $f$, consistent with the prediction of the ST-FMR~\cite{PhysRevB.92.064426}. 
In Fig.~\ref{fig3}(d)--(i), we show in-plane magnetic field angle $\theta_{xy}$ dependence of $S_{\rm{Co}}$ and $A_{\rm{Co}}$, extracted from the $V_{\rm{DC}}$ signals measured at $f$ = 13 GHz, for $\bf{J}$$\parallel [110]$, $\bf{J}$$\parallel [010]$ and $\bf{J}$$\parallel [\bar{1}10]$. In all three devices, $S_{\rm{Co}}$ and $A_{\rm{Co}}$ are proportional to $\rm{sin}2 $$\theta_{xy}$$\rm{cos}$$\theta_{xy}$, consistent with the isotropic SHE with the magnetization orientation in  Fe$_3$Si~\cite{PhysRevLett.131.076702}. Notably, the sign and magnitude of $S_{\rm{Co}}$ and $A_{\rm{Co}}$ are nearly identical in all three samples with $\bf{J}$$\parallel [110]$, $\bf{J}$$\parallel [010]$ and $\bf{J}$$\parallel [\bar{1}10]$. This result suggests that the SHE in Fe$_3$Si is independent of the current orientation.

\begin{figure}[tb]
\center\includegraphics[scale=1]{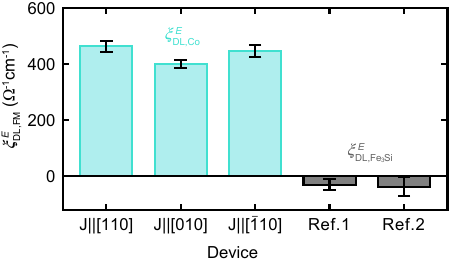}
\caption{
The dampinglike torque efficiency $\xi _{\rm DL, FM}^{\rm E}$ per applied electric field. The green bars are $\xi _{\rm DL, Co}^{\rm E}$ of the Co/Ti/Fe$_3$Si devices with $\bf{J}$$\parallel [110]$, $\bf{J}$$\parallel [010]$ and $\bf{J}$$\parallel [\bar{1}10]$, where the ${\rm Co}$ layer is used as a detecting layer of the SOTs. 
The gray bars are $\xi _{\rm DL, Fe_3Si}^{\rm E}$ of the Ti/Fe$_3$Si device (Ref.1) and Co/Ti/Fe$_3$Si device (Ref.2), where the ${\rm Fe_3Si}$ layers are used as detecting layers of the SOTs.
}
\label{fig5} 
\end{figure}

To quantitatively evaluate the SOT, we calculated the dampinglike torque efficiencies per applied electric field $E$, defined as 
\begin{equation}
\xi_{\mathrm{DL,Co}}^{E}=\frac{2e}{\hbar}\frac{\mu_0M_{\mathrm{s, Co}}d_{\rm{Co}}H_{\mathrm{DL}}}{E}.
\label{xi}
\end{equation} 
 Here, the dampinglike effective field $H_{\rm{DL}}$ can be quantified from the values of $S_{\rm{Co}}$ obtained by fitting the measured $V_{\rm{DC}}$ signals with $\theta_{xy}=45^\circ$, using~\cite{PhysRevB.92.214406, fang2011spin}
\begin{align}
S_{\rm{Co}}=&\frac{I_{\rm{RF}} \Delta R}{2\sqrt{2}} \mu_0H_{\mathrm{DL}} \nonumber \\
&\times\frac{\sqrt{\mu_0H_{\mathrm{FMR, Co}}(\mu_0H_{\mathrm{FMR, Co}}+\mu_0M_{\mathrm{eff, Co}})}}{W_{\rm{Co}}(2\mu_0H_{\mathrm{FMR, Co}}+\mu_0M_{\mathrm{eff, Co}})},
\label{HDL}
\end{align}
where $I_{\rm{RF}}$ is the RF current in the ST-FMR device and $\Delta R$ is the resistance change of the ST-FMR device due to the AMR in the Co layer. 
The contribution from the Fe$_3$Si layer to $\Delta R$ is carefully subtracted by measuring the AMR of a SiO$_2$(4~nm)/Fe$_3$Si(5~nm)/MgO-substrate reference device for each current orientation, resulting in the same value of $\Delta R = 0.62~\Omega$ for all three devices with $\bf{J}$$\parallel [110]$, $\bf{J}$$\parallel [010]$ and $\bf{J}$$\parallel [\bar{1}10]$, consistent with the fact that $\Delta R$ arises from the AMR of the Co layer. 
We determined $I_{\mathrm{RF}}$ by measuring the resistance change of the devices due to the Joule heating induced by applying DC and RF currents~\cite{PhysRevB.92.214406}. 
The result for the Co/Ti/Fe$_3$Si film with $\bf{J}$$\parallel [110]$ (pink), $\bf{J}$$\parallel [010]$ (blue), and $\bf{J}$$\parallel [\bar{1}10]$ (orange) is shown in Fig~\ref{fig4}. In Fig.~\ref{fig4}(a)--(c), we show the resistance change $\Delta R(I_{\mathrm{DC}})= R(I_{\mathrm{DC}})-R(I_{\mathrm{DC}} = 0)$ due to the Joule heating induced by the DC current $I_{\mathrm{DC}}$ application for the devices, where $R(I_{\mathrm{DC}})$ is the resistance of the device under the application of $I_{\mathrm{DC}}$. This result shows that $\Delta R(I_{\mathrm{DC}})$ follows the parabolic relationship to $I_{\mathrm{DC}}$, as expected for the sample heating. We also measured RF power $P$ dependence of the resistance change $\Delta R(P)= R(P)-R(P = 0)$, as shown in Fig.~\ref{fig4}(d)--(f), where $R(P)$ is the resistance of the device under the application of the RF current with the power of $P$. By comparing the resistance changes due to the DC and RF current applications, we obtain the RF current $I_{\mathrm{RF}}$ flowing in the device at each RF power $P$, as shown in Fig.~\ref{fig4}(g)--(i). This result shows $I_\mathrm{RF}\propto \sqrt{P}$, as expected. Using the estimated values of $I_\mathrm{RF}$, we determined $\xi_{\mathrm{DL, Co}}^{E}$ for each current orientation as shown in Fig~\ref{fig5}.

The dampinglike torque efficiency $\xi_{\mathrm{DL,Co}}^{E}$ is dominated by the SHE in the Fe$_3$Si layer. 
In the Co/Ti/Fe$_3$Si(001) device, the SHE in the Ti layer and interfacial effects from the Co/Ti and Ti/Fe$_3$Si interfaces can also contribute to $\xi_{\mathrm{DL, Co}}^{E}$. 
We assume that the contribution from the SHE in the Ti layer to $\xi_{\mathrm{DL, Co}}^{E}$ is negligible because the spin Hall conductivity of Ti is vanishingly small compared to the measured value of $\xi_{\mathrm{DL, Co}}^{E}$~\cite{PhysRevB.90.140407}.  
To estimate the contribution from the Ti/Fe$_3$Si interface to $\xi_{\mathrm{DL, Co}}^{E}$ of the Co/Ti/Fe$_3$Si(001) device, we measured the ST-FMR for a SiO$_2$(4~nm)/Ti(3~nm)/Fe$_3$Si(5~nm)(001)/MgO(001)-substrate reference device, where the Co layer is absent. In this device, the Fe$_3$Si layer is used as a detection layer of the SOTs to evaluate a possible spin current from the Ti/Fe$_3$Si interface.
For the Ti/Fe$_3$Si(001) device, we obtain  
$\xi_{\mathrm{DL, Fe_{3}Si}}^{E} = - 30.6 ~\rm {\Omega}^{-1}\rm cm^{-1}$, which is more than an order of magnitude smaller than the measured value of $\xi_{\mathrm{DL, Co}}^{E}$ for the Co/Ti/Fe$_3$Si(001) device as shown in Fig~\ref{fig5}. This result suggests that the Ti/Fe$_3$Si interface plays a minor role in generating the SOTs. 
For the evaluation of the torque generation at the Co/Ti interface, we also estimated $\xi_{\mathrm{DL, Fe_{3}Si}}^{E}$ for the Co/Ti/Fe$_3$Si(001) device, where the magnetization of the Fe$_3$Si layer detects the SOTs. In this measurement, the SOTs can be generated by the Co/Ti interface, as well as the Ti/Fe$_3$Si interface and the bulk of the Co layer. 
From the measured value of $V_{\rm{DC, Fe_{3}Si}}$ (see Fig.~\ref{fig2}(d), \ref{fig2}(e) and \ref{fig2}(f)), we obtain 
$\xi_{\mathrm{DL, Fe_{3}Si}}^{E} = - 37.3 ~\rm {\Omega}^{-1}\rm cm^{-1}$,
which is more than an order of magnitude smaller than the measured value of $\xi_{\mathrm{DL, Co}}^{E}$ for the Co/Ti/Fe$_3$Si(001) device as shown in Fig~\ref{fig5}. 
Furthermore, the difference in $\xi_{\mathrm{DL, Fe_{3}Si}}^{E}$ between the Ti/Fe$_3$Si(001) and Co/Ti/Fe$_3$Si(001) devices is as small as 
$6.7~\rm {\Omega}^{-1}\rm cm^{-1}$, suggesting the dampinglike torque originating from the Co/Ti interface and the Co layer are negligible. 
Therefore, the dampinglike torque acting on the magnetization of the Co layer is dominated by the SHE in the Fe$_3$Si layer.


The dominant contribution of the SHE in the Fe$_3$Si layer to the dampinglike torque efficiency $\xi_{\mathrm{DL,Co}}^{E}$, shown in Fig.~\ref{fig5}, demonstrates that both the sign and magnitude of the SHE in the Fe$_3$Si are nearly independent of the current orientation. These results indicate that, while the signs of the AMR and PHE vary depending on the current orientation, the signs and magnitudes of the AHE and SHE remain unaffected by the current orientation in epitaxial Fe$_3$Si, as summarized in Table~\ref{table}. This demonstrates the distinct behaviors of the spin-dependent transport phenomena, highlighting that these effects are driven by fundamentally different microscopic mechanisms.

The mechanism for the anisotropic AMR has been investigated in a previous study for single crystal Co$_x$Fe$_{1-x}$ films using a first-principles transport formalism~\cite{PhysRevLett.125.097201}. In this mechanism, the intrinsic AMR is originated from the magnetization-dependent spin-orbit coupling. The spin-orbit coupling can cause the bands to interact and  form an avoided crossing at special $k$ points. In single crystal Co$_x$Fe$_{1-x}$ films, the spin-orbit coupling depends on the direction of the magnetization with regard to the crystalline axis. Scince different magnetization direction gives rise to different spin-orbit coupling, the band crossing/anticrossing also depends on the magnetization direction.
When an avoided crossing is located around the Fermi level, the interband scattering rate increases such that the conductivity becomes smaller. Thus the magnetization dependent band crossing/anticrossing can result in the magnetization dependent conductivity, i.e. AMR. The magnetoresistance due to the magnetization-dependent spin-orbit coupling is also predicted in other single crystal metals, such as Fe and Co~\cite{PhysRevB.108.L020401}.
Here, the AMR arising from this mechanism can depend on the current orientation because the local band structure reflected in conductivity depends on the current orientation.
 For example,  the longitudinal conductivity along [100], $\sigma  ^{[100]} _{xx}$, is determined primarily by the electron velocity along the $k_{[100]}$ direction, which is the gradient of the band energy along $k_{[100]}$ direction. There are also additional contributions from the wave vectors (or $k$ points) outside the $k_{[100]}$ direction, which have nonzero projection on $k_{[100]}$ direction. Nevertheless, the change of $\sigma  ^{[100]} _{xx}$ is expected to most directly reflect the band structure variation along $k_{[100]}$ direction. Similarly, the change of $\sigma  ^{[110]} _{xx}$ is expected to most directly reflect the band structure variation along $k_{[110]}$ direction. Since the band structure along $k_{[100]}$ and $k_{[110]}$ is different from each other, $\sigma  ^{[100]} _{xx}$ and $\sigma  ^{[110]} _{xx}$ show different dependence on the magnetization, and thus AMR can be anisotropic with regard to the current orientation. 
 This scenario can account for our observation of anisotropic AMR in epitaxial Fe$_3$Si films.
Similarly, when the band properties exhibit nontrivial anisotropy depending on the wave vector, the intrinsic PHE associated with the band structure can also become anisotropic, as observed in this study.

The current and magnetization orientation-dependent band crossing/anticrossing may influence the AHE and SHE, as well as the AMR and PHE. This assumption stems from the fact that the avoided crossing near the Fermi energy is associated with an enhancement of the Berry curvature~\cite{RevModPhys.82.1539}, which is the origin of the AHE and SHE. However, our results demonstrate isotropic AHE and SHE. These results align with symmetry arguments; in cubic crystals, the AHE and SHE are isotropic with respect to the current orientation due to the high symmetry constraints of the crystal structure~\cite{PhysRevLett.105.246602}.
One possible explanation for the isotropic AHE and SHE, despite the magnetization-orientation-dependent band crossing/anticrossing, is the cancellation of local effects at special $k$ points when integrated over the entire $k$-space.
The special $k$ points that form band crossings/anticrossings depending on the magnetization orientation can exist at multiple locations in the Brillouin zone. In contrast to the AMR or PHE, which reflects the band structure along a specific direction, the anomalous Hall conductivity or spin Hall conductivity is calculated by integrating the Berry curvature over the entire Brillouin zone~\cite{FENG2016428,RevModPhys.82.1539,PhysRevLett.100.096401}. 
Thus, even if the Berry curvature locally varies with the magnetization, its integration over the entire $k$-space may average out these local variations, resulting in the isotropic AHE and SHE.
In fact, a cancellation of local effects has been observed in the AMR of single-crystal Co$_x$Fe$_{1-x}$ films~\cite{PhysRevLett.125.097201}. For instance, along $k_{[110]}$, there are two special $k$ points with opposite crossing/anticrossing dependencies on the magnetization orientation, resulting in a much smaller AMR amplitude for $\mathbf{J} \parallel [110]$ compared to $\mathbf{J} \parallel [100]$, which has only a single special $k$ point along $k_{[100]}$. This cancellation effect is likely more pronounced for the AHE and SHE, as these phenomena reflect the properties of the entire $k$-space.

\section{IV. CONCLUSIONS}

In summary, we have investigated the spin-dependent transport phenomena in epitaxial Fe$_3$Si films, focusing on the AMR, PHE, AHE, and SHE. The results reveal that while the sign and magnitude of the AMR and PHE depend on the current orientation, those of the AHE and SHE are nearly independent of it. The anisotropic AMR and PHE are attributed to current and magnetization dependent local band properties, including band crossing/anticrossing at specific $k$ points. In contrast, despite the magnetization-orientation-dependent band crossing/anticrossing, the AHE and SHE are isotropic with respect to the current orientation.
This isotropy is attributed to the unique nature of the AHE and SHE, which arise from the Berry curvature integrated over the entire Brillouin zone, effectively canceling local variations caused by magnetization-dependent spin-orbit coupling in $k$-space. These findings highlight the distinct microscopic mechanisms underlying these spin-dependent phenomena: while AMR and PHE are influenced by the local band structure along specific directions, AHE and SHE are governed by global $k$-space properties, reflecting the interplay between symmetry, band structure, and magnetization.

\begin{acknowledgments}
\section{ACKNOWLEDGMENTS}
This work was supported by JSPS KAKENHI (Grant Number: 22H04964, 24H00034), Spintronics Research Network of Japan (Spin-RNJ), and MEXT Initiative to Establish Next-generation Novel Integrated Circuits Centers (X-NICS) (Grant Number: JPJ011438). N.S. is supported by JSPS Grant-in-Aid for Research Fellowship for Young Scientists (DC1) (Grant Number 22KJ2714), and Keio University Doctorate Student Grant-in-Aid Program from Ushioda Memorial Fund. 
\end{acknowledgments}

%

\end{document}